\documentclass[12pt,preprint]{aastex}

\begin{document}
\title{Chemical Enrichment in Damped Lyman $\alpha$ Systems
From Hierarchical Galaxy Formation Models}
\author{Katsuya Okoshi, Masahiro Nagashima\altaffilmark{1} and Naoteru Gouda}
\affil{National Astronomical Observatory,
Mitaka,  Tokyo 181-8588,  Japan;
\email{okoshi@th.nao.ac.jp}}
\and
\author{Satoshi Yoshioka}
\affil{Department of Physics, Tokyo University of Mercantile Marine,
Tokyo 135-8533,  Japan}

\altaffiltext{1}{Present address: Department of Physics, University of
Durham, South Road, Durham DH1 3LE, United Kingdam}
\begin{abstract}
We investigate chemical enrichment in Damped Lyman $\alpha$ (DLA)
systems in the hierarchical structure formation scenario using a
semi-analytic model of galaxy formation.  The model developed by
Nagashima, Totani, Gouda \& Yoshii takes into account various selection
effects on high-redshift galaxies and can show fundamental
observational properties of galaxies, such as luminosity functions and
number-magnitude/redshift relations.  
DLA systems offer the possibilities of 
measuring metal abundance more accurately than faint galaxies. 
For example, recent measurements of zinc
abundance can provide good evidence for understanding the processes of
metal pollution and star formation in DLA systems because zinc is
 virtually unaffected by dust depletion.  Here we focus on this advantage for
observation in order to explore the metallicity evolution in DLA systems
at high redshifts.  
We can consistently show the metallicity evolution for reasonable models 
which also reproduce fundamental properties of local galaxy population.
This result suggests that the chemical evolution of DLA
systems can be consistently reconciled with the observational features
of typical galaxies.  We also investigate other properties of DLA
systems (column density distribution and mass density of cold gas), 
and find that star formation in massive galaxies should be
more active than that in low-mass ones. This is consistent with 
the results by Nagashima et al. and Cole et al. in which 
the star formation timescale is set by reproducing 
cold gas mass fraction in local spiral galaxies.
Finally we discuss host galaxies associated with DLA systems. 
We conclude that they primarily consist of sub-$L^{*}$ and/or dwarf 
galaxies from the observations. 
\end{abstract}



\section{Introduction} 

The absorption-line systems observed in the spectra of background
quasars have been proven to be a strong probe into the physical conditions
of the Universe at high redshifts: 
abundances of neutral gas and metals, kinematic properties, etc. 
In particular, Damped Lyman $\alpha$ Absorption (DLA) systems, 
which have been associated
with cold gas in protogalactic disks \citep{WTSC}, provide some advantages
in the investigation of various characteristics of primordial galaxies similar to 
faint ones at high redshifts. 
For example, we can obtain abundant data in the early universe from 
quasar spectra which are less affected by observational limits 
caused by faintness of absorbers and chemical abundance, etc. 
Therefore, they have been studied extensively at
redshifts $0\la z \la 4.5$ to explore the physical processes in
galaxies: star formation rates, abundances of metals and dust
\citep{FP93, L96, PSKH, PESB, P00, PW99, PW00}, column density
distribution \citep[e.g.][]{SW00, RT00}
and kinematics \citep{PW97, PW98}. 
These observational features would provide important
constraints on clarifying the evolutionary link between DLA systems
and typical galaxies. 

In particular, in DLA systems, we can measure
abundances of various elements (Fe, Si, Ni, Mn, Cr, Zn, etc.) and
thereby find good clues to the metallicity evolution of galaxies
in the early universe.  For example, relative elemental abundances are
one of the most important complements to this information because abundance
ratios of elements produced by supernovae in different proportions can
provide independent insights into the chemical evolution.
However, it has been difficult to establish precisely some abundance
patterns (e.g. [$\alpha$/Fe]) in DLA systems because 
the level of depletion of refractory elements onto dust grains 
may be uncertain. 
For this reason, the zinc
abundance has recently been recognized as a good tracer of the metallicity
evolution instead of \ion{Fe}{2} because zinc is one of the undepleted
elements.  Recent observations have generally reported that the absolute
Zn-abundance is $\sim 10 \%$ of the solar value with no clear evolution
over the redshift \citep{PSKH, PESB, PW99, VBCM}.  The low abundance
and the apparent lack of redshift evolution have been considered 
interesting problems for DLA systems as proto-galaxies, in particular, at
low redshifts where abundances much closer to the solar ones are
expected.  This advantage of Zn-measurement confirms that DLA systems
so far remain one of the best probes at high redshifts into star
formation history.

Recent theoretical studies focus on the nature of DLA systems and 
physical relations with galaxies observed at high and low redshifts. 
\citet{Ka96} studied the formation of DLA and Lyman-limit systems in 
a hierarchical clustering scenario using a hydrodynamical simulations 
of a CDM universe($\Omega=1$). They concluded that the
 observed 
\ion{H}{1} column  density distributions can be reproduced within a 
factor of 2 if DLA systems are relatively massive and dense
protogalaxies. They also showed that their model could reproduce 
the other properties of DLA observations. However, these results remain 
somewhat inconclusive. The calculations were restricted to a single CDM 
cosmology ($\Omega=1$ CDM). In addition, they could not include galactic 
disks embedded in halos with circular velocities 
$V_{\rm c} \la 100$ km s$^{-1}$ caused by limited resolutions 
($M \sim 10^{11}~h^{-1}~M_{\odot}$). 
Gardner et al.(1997a,b, 2001) resolved these problems by using 
a relation between absorption cross-section $\sigma$ and 
halo circular velocity $V_{\rm c}$ obtained from numerical simulations. 
Assuming that the $\sigma - V_{\rm c}$ relation can be applied for 
previously unresolved halos, they presented 
numerical results for various cosmological models. As a result, the
observed abundance of DLA systems can be reproduced if their
extrapolation procedure applied for less-massive halos 
($V_{c} \ga 50-80$ km s$^{-1}$). However, the validity of introducing 
their correction method remains uncertain. Therefore, their calculations 
could not still include the contribution of DLA systems below limited 
resolutions, especially at redshifts $0 \le z \le 3$. 
Haehnelt, Steinmetz \& Rauch (1998, 2000) focussed on the kinematics 
to explore the nature of DLA systems in SPH simulations with high
resolutions ($\sim$ a few kpc). 
They concluded that the high abundance of protogalactic clumps can 
reproduce the observed velocity width distribution and asymmetries of 
absorption lines found by Prochaska and Wolfe (1997,1998). 
Their conclusion suggested that the {\it majority} of DLA systems are not 
large, rapidly rotating disks with $V_{\rm c} \ga 100$ km s$^{-1}$ 
but protogalactic clumps with the typical circular velocity of 
DLA halos $V_{\rm c} \sim 100$ km s$^{-1}$. 
Their numerical simulations presented that their models can reproduce 
observational properties of DLA systems and provided interesting
insights to explore the nature of DLA systems.

Recently a different approach, semi-analytic modelling, has been 
applied with a view to deciphering the clues to the formation 
process of galaxies in the hierarchical clustering scenario 
(e.g., Kauffmann, White, \& Guiderdoni
1993; Cole et al. 1994, 2000; Somerville \& Primack 1999; Nagashima,
Gouda \& Sugiura 1999; Nagashima et al. 2001, hereafter NTGY; 
Nagashima et al. 2002).  
These approaches have some advantages. For example, 
semi-analytic modelling can study the effect of star
formation or supernovae feedback on galaxy evolution 
even under simple recipes. It does not also suffer from 
resolution limitations. This approach can also take 
into account merging the histories of dark halos based on 
the power spectrum of the initial density fluctuation, and 
have successfully provided galaxy formation models for 
reproducing observational properties of galaxies such as 
luminosity functions, the Tully-Fisher relation, the relation 
between \ion{H}{1} gas mass fraction and luminosities, and so forth.
To understand the galaxy formation process, it is also important to
focus on observational features of high redshift galaxies.  For example,
the faint galaxy number counts are good evidence for
constraining the key process of galaxy formation in addition to the
local feature.  NTGY confirms, using a similar semi-analytic model, that
the fundamental properties of local galaxies can be reproduced by their
semi-analytic galaxy formation model in a cosmological
constant-dominated flat universe ($\Lambda$CDM model). 
Their model shows good agreement with the observational features containing those of
galaxies at high redshifts such as galaxy counts, taking into account
high-redshift selection effects caused by cosmological dimming of
surface brightness, absorption by intergalactic \ion{H}{1} gas, and
internal dust absorption.  Therefore, it is valuable to examine how 
improvements in the above model affect our predictions for the
metallicity evolution of cold gas in high-redshift galaxies.  
We investigate here the star formation history of DLA systems in our
semi-analytic model, 
and whether we can consistently reproduce the metallicity evolution 
for the models which best match the observational constraints on 
local galaxy population. 
We also address what kinds of physical
process, such as star formation timescale, play key roles in the
evolution of DLA systems within the framework of the semi-analytic approach
investigated here. 
Furthermore, we study other properties of DLA systems: \ion{H}{1} column 
density distribution, mass density of cold gas and so on. 
Finally we discuss what kinds of galaxies are associated with DLA
systems. 

In \S 2, we briefly describe the semi-analytic model used here.  In \S
3, we show the evolution in chemical enrichment of DLA systems. 
In \S 4, we present other properties of DLA systems in our calculation. 
Host galaxies of DLA systems are also discussed in this section. 
Finally we summarize our conclusions in \S 5.


\section{Model}
We use the semi-analytic model for galaxy formation based upon the work
in NTGY.  As shown in NTGY, our model reproduces luminosity
functions, cold gas mass fraction, and disk radius of galaxies in the
local universe and the faint galaxy number counts in the $\Lambda$CDM
model.

For cosmological parameters, we adopt $\Omega_{0}=0.3$,
$\Omega_{\Lambda}=0.7$, $\Omega_{\rm b}=0.015h^{-2}$, $h=0.7$ 
(where $h$ is the Hubble parameter, $h=H_{0}/100$km~s$^{-1}$
Mpc$^{-1}$), and
$\sigma_{8}=1$ that is the normalization of the power spectrum
of density fluctuation given by \citet{BBKS}.  The cosmological
parameters adopted here are consistent with recent observations, 
e.g., WMAP  \citep{Spergel03} while a slightly higher $\Omega_{\rm b}$ is 
favoured recently. In this paper we generally follow the model
examined in NTGY.  In the following, we briefly describe aspects of
the model.

The number density of progenitors of a dark halo as a function of their
mass and redshift is given by an extended Press-Schechter model
\citep{PS74, BCEK, B91, LC93}. The number of local halos follows 
the Press-Schechter mass function. The merging process of dark halos is
given by the method developed by \citet{SK99} based on a Monte Carlo
method.  We focus on halos with circular velocity $V_{\rm circ}
\geq 40$ km~s$^{-1}$ and treat systems with small $V_{\rm circ}<40$
km~s$^{-1}$ as diffuse accretion mass.

We assume that baryonic gas consists of two phases: cold and hot.  When
a halo collapses, halo gas is assumed to be shock-heated to the virial
temperature of the halo and distributed in a singular isothermal sphere
({\it hot} gas).  The cold gas is defined as the gas component
within a `cooling' radius in which hot gas cools quickly. 
Only a `central' galaxy in a halo accretes the cold
gas.  The cold gas then becomes available for star formation.  The star
formation process can play an important role in chemical enrichment of 
cold gas.  The star formation rate (SFR) is assumed as
\begin{eqnarray}
\dot{M}_{*}=\frac{\displaystyle M_{\rm cold}}{\displaystyle 
\tau_{*}}, 
\end{eqnarray}
where $M_{*}$ and $M_{\rm cold}$ are the mass in stars and cold gas,
respectively, and $\tau_{*}$ is the timescale of star formation.
Because the star formation process has large uncertainties, it is
conceivable that SFR takes place in different ways.  
In our model, we
assume two types of star formation timescale: constant star formation
(CSF) and dynamical star formation (DSF)as follows,
\begin{eqnarray}
\tau_{*}=\left\{
\begin{array}{ll}
\displaystyle{\tau_{*}^{0}\left(\frac{V_{\rm circ}}{V_{*}}\right)
^{\alpha_{*}}} & \mbox{(CSF)},\\
\displaystyle{\tau_{*}^{0}\left(\frac{V_{\rm circ}}{V_{*}}\right)
^{\alpha_{*}}\left[\frac{\tau_{\rm dyn}(z)}{\tau_{\rm dyn}(0)}\right]
} & \mbox{(DSF)}.
\end{array}
\right.
\end{eqnarray}
Note that these two are
correspond the constant efficiency (CE) and accelerated efficiency
(AE) models, respectively, in \citet{spf01}.  For the CSF model, the star
formation timescale is constant at all redshifts $z$.  For the DSF model,
the timescale is proportional to the dynamical time of disks, $\tau_{*}
\propto \tau_{\rm dyn}(z)$, which becomes shorter as redshift
increases. According to \citet{CAFNZ, CLBF}
and NTGY, we also take into account the dependence of the star formation
timescale on the circular velocity, $\tau_{*}\propto V_{\rm
circ}^{\alpha_{*}}$. 
We here adopt LC and LD models in NTGY as reference models in which 
the star formation timescales are given by 
$(\tau_{*}^{0}, V_{*})=(1.5\mbox{Gyr}, 300$ km~s$^{-1}$) for CSF 
and $(4\mbox{Gyr}, 200$ km~s$^{-1}$) for DSF, respectively.  
Actually these parameters are determined by matching the
cold gas mass fraction of spiral galaxies to the one observed.  Therefore,
these are also expected to play an important role in determining the
observable characteristics of DLA systems.
It should be also noted that we here explore two types 
of star formation timescale according to the notation defined by
\citet{sp99}. One is `Durham model' in which the star formation
timescale depends only on circular velocity \citep{CAFNZ} and another is
`Munich model' in which it depends only on redshift \citep{KWG93}.
Thus the star formation timescale is here 
assumed to totally depend on the circular velocity of halos and
redshifts. 
For chemical enrichment, 
we take metal yield $y=0.038$ adopted in NTGY. 

Supernova explosions also affect galaxy evolution.  We follow here a
recipe for the supernova feedback process formulated in NTGY: the
reheating rate of cold gas by supernova explosions is assumed to be
proportional to SFR,
\begin{eqnarray}
\dot{M}_{\rm reheat}=\left(\frac{\displaystyle V_{\rm circ}}
{\displaystyle V_{\rm hot}}\right)^{- \alpha_{\rm hot}}\dot{M_{*}}, 
\end{eqnarray}
where the values of parameters $V_{\rm hot}=280$ km~s$^{-1}$ and
$\alpha_{\rm hot}=2.5$ are required by reproducing the observed
luminosity function of local galaxies in LC and LD models in NTGY.
Note that the conclusions in the previous work show that the supernova
feedback process is tightly coupled with the shape and normalization of 
the luminosity function of local galaxies.

When two or more halos merge, the central galaxy in the largest
progenitor halo becomes the new central one.  All other galaxies, called
`satellite galaxies', remain to be placed around the center.  After each
halo-merging, we consider the following two mechanisms for mergers of
galaxies: dynamical friction and random collision.  Satellite galaxies
merge with the central one in the dynamical friction timescale.
Between satellite galaxies, random collisions also occur in the mean free
timescale.  When two galaxies merge, if the mass ratio of the smaller
galaxy to the larger one is larger than $f_{\rm bulge}$, a starburst
occurs and all cold gas is consumed(major merger).  Then all stars compose a
spheroidal component.  Otherwise, the smaller one is simply absorbed
into the disk of the larger one with no additional star formation
activity(minor merger).
We summarize parameter sets adopted for LC and LD models in Table 1 
(the other parameters are the same as in NTGY).  

Finally we define DLA systems in our model. 
We simply assume that all DLA
systems have gaseous disks which are face-on to us (the inclination 
effect will be discussed in section 4) and 
the radial distribution of \ion{H}{1} 
column density follows an exponential profile, $N_{\rm HI}(r)=N_{\rm 0} 
\exp(-r/r_{\rm e})$, where $N_{\rm 0}$ is the central column density 
of neutral gas and $r_{\rm e}$ is the effective radius of gaseous
disks. We here assume the effective radius 
$r_{\rm e}=r_{0} (1+z)$ where $r_{0}$ is the radius provided by 
specific angular momentum conservation of cooling hot gas\citep{NYTG}.
According to the paper, the specific angular
momentum of halos has a log-normal distribution in terms of the
so-called nondimensional spin parameter $\lambda$ but with a slightly
large mean and dispersion, $\bar{\lambda}=0.06$ and 
$\sigma_{\lambda}=0.6$.
The central column density $N_{\rm 0}$ is given by 
$N_{\rm 0}=M_{\rm cold}/(2 \pi \mu m_{\rm H} r_{\rm e}^{2})$,
where $m_{\rm H}$ is the mass of a hydrogen atom and $\mu(=1.3)$ is the
mean molecular weight. 
The size of DLA systems is defined by the radius $R$ at which 
$N_{\rm HI} = 10^{20}$ cm$^{-2}$. 
For each system, we take the column density averaged over radius within 
$R$. 

\section{Chemical Enrichment in  DLA Systems} 

Figure 1 shows the metallicity evolution of cold gas in DLA systems.
The solid line shows a result for the LC model ($\tau_{*}^{0}=1.5$ Gyr
and $\alpha_{*}=-2$). 
We compare our results here with the data given by
\citet{PW00}   
\footnote{The observed metallicity [Fe/H] from \citet{PW00} 
is here corrected to [Zn/H] adopting [Zn/Fe]=0.4. Here we present 
[Zn/H] as observational data} and \citet{S00}. 
\citet{S00} evaluates the metallicities of DLA systems adopting dust
correction and shows ranges of the metallicity as a function of
redshifts  ($0 \la z \la 4.5$).  
Our calculations show that the mean metallicity increases
gradually from $1/30$ of solar abundance at redshift $z \sim 3$ to a 
half at the present due to ongoing star formation with
constant efficiency.  The metallicity evolution shows good agreement
with the observations in the whole range of redshifts. 
Thus the LC model can reproduce the metallicity evolution
of DLA systems, as well as many galactic properties such as luminosity 
function and galaxy counts. 
Note that the dotted line indicating mean stellar metallicity 
nearly traces that of cold gas. This is because cold gas is the 
direct material of stars. 

The efficiency of star formation plays a key role in the
metallicity evolution in DLA systems.  In Figure 1, we also present a  
result for the LD model (dashed line) 
in which the star formation timescale is 
proportional to the dynamical time of disks, $i.e.$, 
SFR is higher at high redshifts.
The resultant curve for the LD model is very shallow (1/10 solar
abundance at $z\sim 5$) because SFR is much larger at high redshifts
compared to that in the LC model.  Therefore this model results in
somewhat over-abundance compared to the high-redshift data.  
We thus conclude that the star formation timescale should be nearly constant in
order to obtain a good agreement with the observed data for both DLA
systems and local galaxies. 

Now we turn to exploring the dependence of the chemical enrichment 
on the star formation process around the best model to reproduce the 
cold gas fraction well in local spiral galaxies. 
Figures 2(a) and 2(b) depict the effects of changing
$\tau_{*}^{0}$ and $\alpha_{*}$, respectively, on chemical enrichment.
The metal abundance strongly depends on $\tau_{*}^{0}$ and $\alpha_{*}$ as
well as on the cold gas mass fraction of galaxies 
(see Figure 9 in \citet{CLBF}). 
Figure 2(a) shows the dependence on $\tau_{*}^{0}$ in the LC model.  
As the star formation timescale is shorter (SFR is larger), 
the average metallicity becomes higher through metal enrichment
process. We also show the dependence on $\alpha_{*}$ on the 
metallicity in Figure 2(b). In the LC model with $\alpha_{*}= -2$,
comparing to the other model with $\alpha_{*}=0$ that roughly
corresponds to `quiescent model' in \citet{spf01}, 
DLA systems have low metallicities ($\sim 1/10 Z_{\odot}$)
at high redshifts. This is because the negative value of 
$\alpha_{*}$ decreases SFR in low mass galaxies, which are numerous 
and has generally low metallicity, and then leads them to be identified as
DLA systems due to increasing gas frcation. 
Therefore, our results are different from those found in \citet{spf01} 
in which DLA systems have higher metallicities of
cold gas than observational data at high redshifts.
We also checked the
dependences of our results on other parameters such as the SN
feedback-related parameters $V_{\rm hot}$ and $\alpha_{\rm hot}$ and
found that they only weakly affect metallicity
evolution in the whole range of redshifts.
Thus, from these figures, we find that the metallicity evolution of DLA
systems can be consistently reproduced by the LC model. 
We will see that the LC model can also successfully reproduce the column 
density distribution in the next section. 

Next, in Figure 2(c), we present results not only for LC but for the
the standard CDM (SC) and an open CDM (OC). 
Astrophysical parameters for SC and OC models are normalized in the 
same manner as LC and LD models determined by local galaxy properties
(Table 1).  
The baryon density parameters $\Omega_{\rm
b}=0.015 h^{-2}$ are adopted for all models.  The value of $\sigma_{8}$
is normalized by cluster abundance for models LC and OC.  Note that NTGY
found that LCDM is favoured by the observed HDF galaxy counts.  
Figure 2(c) shows that the LC model predicts the gas metallicity in the best 
agreement with the observed data among these cosmological models at all
redshifts.  While both SC and OC models totally underpredict the observed
metallicity, the LC model exhibits milder evolution at high redshifts compared
to the SC model and show a good agreement with the observations.  Thus, also
from the metallicity evolution of DLA systems, low density universes are
favoured.

\section{Global Properties of DLA Systems} 

In this section, we focus on \ion{H}{1} column density distribution 
of DLA systems because we need to confirm 
that gas systems selected here are DLA systems. 
In Figure 3, we show differential distribution 
$f(N_{\rm HI},X(z))=d^{2}N / dN_{\rm HI} d X$, 
which denotes the number per unit column density $N_{\rm HI}$ and per unit absorption 
distance $X(z)$ \citep{BP69}. 
The data points are taken from \citet{SW00}
\footnote{We adjust the data points for the cosmological model(LCDM)
adopted here(Table 1).} . 
This result is averaged over redshifts $0 \le z \le 5$. 
The solid line shows a result for the LC model.  
The column density distribution is entirely in good agreement with 
the observational data.   
Alternatively we also present a 
result for the LD model (dashed line)in Figure 3. 
In this case, DLA systems are totally deficient over the whole
range of column density. The LD model has a shorter timescale 
of star formation by which cold gas turns into stars more 
rapidly than the LC model at high redshift. Therefore, the LD
model predicts smaller number of DLA systems than the LC model.
Thus, the efficiency of star formation also plays an important 
role in the column density distribution. 
 
It is also interesting to investigate the inclination of
galactic disk. The inclination averaged cross section is half
as much for random orientations although we here assume that
all disks are face on to us. This corresponds to $\theta=60^{\circ}$ 
when a disk is viewed at an angle $\theta$ to the normal.
We therefore calculate the column density distribution for
viewed angles: $\theta=0^{\circ},~30^{\circ}$, and $60^{\circ}$. 
Figure 4 shows how the differential distribution depends on the inclination. 
We find that the slope in the distribution is a little
flatter when the viewed angle becomes larger because the cross section
becomes small and the averaged $N_{\rm HI}$ increases by the
inclination. However, the differences between the results
are over all small ($\la 0.1$ dex) in the observed range of column
density ($N_{\rm HI} > 10^{20.3}$ cm$^{-2}$). 
Therefore, in this paper, we assume face-on for all galaxies because of 
the small dependence on the inclination. 

To investigate the number evolution of DLA systems, in Figure 5, 
we also show differential distribution at four redshifts: 
(a)$z=1$, (b)$z=2$, (c)$z=3$, and (d)$z=4$. 
The observational data are from \citet{SW00}. 
In Figure 5(a)(open circles), we also take additional data reported by
\citet{RT00} which include a sample at low redshift 
$ \langle z \rangle \sim 0.8$. 
We find that our results are generally consistent 
with the observations.   
\citet{SW00} and \citet{RT00} also pointed out that the slope in  differential 
distributions is more moderate at lower redshifts.
This is an interesting signature to discuss concerning the formation of DLA
systems.  
However, it is still controversial to address some statements 
about the redshift evolution of $f(N_{\rm HI},X)$ systematically 
because the samples are not so large.
While the LC model can provide a reasonable fit to 
observations, low $N_{\rm HI}$ systems are slightly deficient at 
redshift $z=4$. 
Although we need more observational samples to see whether this deficient should be
serious, this might require some mechanism enlarging the cross section 
such as galactic winds pushing gas out far away from central disks.

Previously, \citet{MPSP} investigated radial distribution 
of cold gas in DLA systems at high redshifts using a semi-analytic model
given by \citet{spf01}. They found that 
a Mestel distribution can reproduce column density 
distributions better than single-disk models (also including 
exponential distribution) based on angular momentum conservation. 
Although \citet{MPSP} required number suppression of DLA 
systems in order to match the kinematic data,
the single-disk models fail to reproduce the column density distributions 
because DLA systems have a short radial extent so that 
the cross sections are too small. 
Figure 3 also shows results for the same model as the LC model 
but for $\alpha_{*}=0$(dotted line). 
This model roughly corresponds to one in which the radial  
distribution of column density is of an exponential type in 
 \citet[][see their Figure 4]{MPSP}. 
Compared with this result, our LC model predicts that 
DLA systems are more abundant, especially for low $N_{\rm HI}$ 
systems, because the negative $\alpha_{*}$ suppressed star formation 
in low $V_{\rm circ}$ systems (see below Figure 8). 
Therefore the number density is consistently high enough to 
reproduce column density distributions of DLA systems even in which 
the radial distribution of \ion{H}{1} column density follows an exponential 
profile.  

Generally, in Figures 3 and 5, we conclude that 
the LC model in our calculation 
can reproduce fundamental properties of DLA systems: 
both the metallicity evolution and the column density distribution, 
so that we investigate 
other properties of DLA systems in the LC model as we discuss below. 

Figure 6 shows the mass density in HI gas contributed by DLA systems. 
The observed data are given by \citet{SW00} and \citet{RT00}. 
Our results show that the evolution of mass density is generally 
similar to that calculated in observations. 
However, the calculated density is higher than the 
observed data at $z \la 2.5$.
As discussed above, the metallicity is generally higher at low
redshifts(Figure 1). 
This might stem from the fact that dusty systems are abundant 
at the present and systematically fail to be identified as DLA systems. 
Therefore, the local observations only show the lower limits of mass 
density in cold-gas systems. 
Furthermore, mass density is very sensitive to the number of 
high-$N_{\rm HI}$ systems if the power-law index $\beta$ ($f \propto
N_{\rm HI}^{- \beta}$) is less than 2.
At $z \la 2.5$, the small-number statistics of high-$N_{\rm HI}$
systems makes assessment of the mass density more uncertain, 
as differences in error bars of observational data between 
low and high redshifts in Figure 6 show.
The larger sample at $z \la 2.5$ would determine the data 
with sufficiently small errors to permit 
comparison with the high-redshift data. 

Figure 7 shows average mass evolution of each phase.
While cold gas mass is almost constant against redshift, 
stellar mass increases gradually towards the present because 
star formation proceeds to accumulate stellar mass  
and the merging process continues to form massive systems.  
At $z \sim 1$, 
the stellar mass exceeds that of cold gas and finally attains 
$10$ times of cold gas in mass at present. 
We also calculate mass-weighted metallicities of cold gas in both
DLA systems and all galaxies. The results are shown in Figure 7. 
Similarly to the evolution of metal mass, the metallicity 
increases gradually towards the present. Futhermore, the
metallicities of cold gas in both DLA systems and all galaxies
are quite close each other. This indicates the fact that the
chemical enrichment in DLA systems are similar to that in all 
galaxies.

Finally, we focus on what kinds of systems are identified as 
DLA systems. Figure 8 depicts average circular velocity of 
DLA systems at $0 \le z \le 5$.
Similar to the mass evolution, the average circular velocity increases 
towards low redshifts as merging proceeds, and typical 
velocities are 
$V_{\rm circ} \sim 60$ km s$^{-1}$ at $z \sim 3$ and  
$V_{\rm circ} \sim 90$ km s$^{-1}$ at $z \sim 0$. 
Figure 9 shows the distribution of circular velocity. Here we
present number fraction of galaxies identified as DLA systems as
a function of circular velocity at redshifts (a)$z=0$, (b)$z=1$,
(c)$z=3$, and (d)$z=5$, respectively. From these results, it is
apparent that the dispersion of $V_{\rm c}$-distribution increases 
gradually towards low redshifts, similarly to the average of 
circular velocity. This implies that massive DLA systems 
form predominantly through galaxy mergers at low redshifts. 
At present, a few percents ($\sim 5-10 \%$) of DLA systems can be 
massive systems like our Galaxy,
while gaseous disks like Milky Way rarely give rise to DLA systems
at high redshifts $z \ga 3$.
This result indicates that DLA systems are more likely to be found in 
less massive halos 
compared with typical $L^{*}$ spirals. 
We find that this picture is strong conflict with the classical one 
in which DLA systems are relatively massive galaxies like our Galaxy. 

Figure 10 shows absolute luminosity evolution (in $U$, $B$, $V$ and $K$ bands) of 
median host galaxies identified as DLA systems. Similarly to stellar mass, 
the luminosity gradually increases with the star formation. 
At present, the average luminosity attains 
$L(B) \sim 2 \times 10^{9} L_{\odot}(B)$ and 
$L(K) \sim  \times 10^{10} L_{\odot}(K)$ 
comparable with dwarf galaxies. 
Considering the other results from our calculation, 
{\it we find that host galaxies of DLA systems primarily 
consist of sub-$L^{*}$ and/or dwarf galaxies. } 
Recently, \citet{COPW} investigated metallicity evolution of DLA 
systems utilizing hydrodynamic simulation. They found that DLA systems 
in the simulation provide good matches to 
observational data concerning metallicity evolution, column density distribution, 
redshift evolution of the neutral gas content, and so on. 
Simultaneously they also stressed that DLA systems comprise 
a mix of various morphological types including less massive systems than 
present-day $L^{*}$ galaxies, and show that the median DLA system typically 
shows absolute luminosity: $L=0.1 L^{*}(z=0)$ at $z=3$ and 
 $L=0.5 L^{*}(z=0)$ at $z=0$. This shows good agreement with 
our result, and also suggests that DLA systems are primarily composed
of faint galaxies. 
Moreover, HST and ground-based observations demonstrate the direct
imaging of DLA systems at $z \la 1$. 
The results suggest that DLA systems might have mixed types of galaxies 
and a number of the systems are dwarf galaxies 
and/or compact objects \citep[e.g.][]{RT98}.    

So far, the nature of DLA systems still remains controversial. 
In principle, DLA systems have
a large column density in neutral hydrogen $N_{\rm HI} \ga 10^{20}$ 
cm$^{-2}$ comparable to the surface density of 
present-day spiral galaxies. This suggests a possibility that 
DLA systems arise from galactic large disks \citep{WTSC}. 
Moreover, Prochaska and Wolfe (1997, 1998) found that absorption
lines of low-ionization ionic species in DLA systems show large velocity 
spreads and partly asymmetric profiles. They also argued that the
observed kinematics can be reproduced by massive rotating disks. 
Alternatively, \citet{HSR98} showed that the majority of DLA systems 
are protogalactic clumps (typical circular velocity $V_{\rm c} \sim 100$ 
km s$^{-1}$) utilizing hydrodynamic simulations within the hierarchical
structure formation scenario. They also stressed that the asymmetries
and large velocity spreads of absorption lines associated with DLA
systems can be reproduced by complex geometry and nonequilibrium
dynamics of neutral gas embedded in dark halos that are not necessarily 
virialized. 
In the present work, we focus on the virialized systems, 
the effects of nonequilibrium dynamics of clumps are neglected, which 
might be important. 
Then further investigations will be required to clarify the kinematics 
of DLA systems in detail. Besides significantly larger sample 
than those at present, $\sim 50$,  should be needed to discuss 
the kinematic data of DLA systems statistically. 
  
Figures 11 show the cross-sections $\sigma_{\rm DLA}$ as a function of
circular velocity $V_{\rm c}$ at redshifts (a)$z=0$, (b)$1$, (c)$3$ 
and (d)$5$. 
This result suggests that DLA systems have typical size 
$\sim 10$ kpc at $V_{\rm c} \sim 200$ km s$^{-1}$. 
When these relations are fitted by power-law, 
$\sigma_{\rm DLA} \propto V_{\rm c}^{\beta}$, we find that 
$\beta \sim 2-3$ for each panel and that $\beta$ increases slightly 
as redshift decreases. This relation has been investigated in 
hydrodynamic simulations. 
For example, \citet{G01} showed steep scaling $\beta \sim 1.5$ for 
a LCDM model. Recent SPH simulations similarly found $\beta \sim 2-3$ 
\citep{HSR00,NSH03}. 
While the steepness of $\sigma_{\rm DLA}-V_{\rm c}$ relations 
in our model is similar to that in numerical simulations, 
the average cross-section in our results is smaller than the numerical ones. 
Such large cross-sections in simulations may arise from two reasons. 
First, the limited resolution prevents us from resolving DLA systems 
with large cross-sections because they have too small cross-sections 
below numerical resolutions.
Practically, some simulations predicted that DLA systems have quite 
large disks, $e.g.\sim 100$ kpc at $V_{\rm c} \sim 200$ km s$^{-1}$.  
Second, the large cross-section can be produced by neutral gas with 
complex geometry or/and nonequilibrium dynamics induced by 
frequent merging of protogalactic clumps. 
It is also produced even by numerical simulation with high resolution 
while the average cross-section is small, $\sim 17$ kpc at 
$V_{\rm c} \sim 200$ km s$^{-1}$ \citep{HSR00}.
Therefore, the different sizes from numerical simulations may mainly 
stem from neglecting the latter effect.

\section{Conclusions}
We investigated the metallicity evolution of DLA systems in a
hierarchical galaxy formation scenario using a semi-analytic galaxy
formation model given by NTGY, which has been found to show 
good agreement with many properties of galaxies.  In the previous 
theoretical work, it has been claimed that the metallicity of DLA 
systems is too high to reproduce the observed ones
\citep[e.g.][]{spf01}.  We find that, in 
contrast to previous work, DLA systems in our model have low
metallicity, $\sim 1/10 Z_{\odot}$, consistent with observational data 
when we use the same model as that reproducing many observations of 
local and high redshift galaxies given by NTGY.  
This conclusion is given by setting the star formation timescale as follows:
(1) nearly constant against redshift, and (2) longer timescale in lower
circular velocity systems.  Our results suggest $\tau_{*} \propto V_{\rm
circ}^{\alpha_{*}}$ and $\alpha_{*}=-2$, which is entirely consistent with NTGY.
The first assumption (1) leads that there is a large amount of
cold gas at high redshift. The second (2) indicates 
that a stronger $\alpha_{*}$ dependence causes low star formation
rate in low mass galaxies that contribute more to the total number
of DLA system. These points
mainly led to the result that DLA systems have low metallicity
($Z \sim 1/10 Z_{\odot}$) consistent with observational data. 

We can also reproduce column density distributions 
{\it even} under the assumption that 
the radial distribution of \ion{H}{1} column density follows
the exponential profile. 
In addition with our results for the average circular velocities and 
absolute luminosities, our calculation indicates that DLA systems are 
primarily composed of less massive systems rather than present-day 
$L^{*}$ galaxies. 
In a subsequent paper, we intend to investigate in detail host galaxies 
of DLA systems in our model, including comparison with observational 
properties of dwarf galaxies and/or the other types of galaxies. 

In further analyses, the following effects might be considered:  First,
in our model, the metal enrichment of cold gas is tightly coupled with
star formation because we assume here that the metal ejected by stars is
completely mixed with cold gas and finally becomes distributed
throughout the disk.  However, it is preferable to consider that galactic disks
generally have metallicity gradients: outer regions may be 
metal-poor contrary to the metal-rich inner regions \citep{P97, T98}.
Second, the metal distribution might not be consistently diffuse.  \citet{CK00}
pointed out that typical DLA systems have a multi-phase medium composed of
cold and warm \ion{H}{1} gas induced by the observational fact that the
spin temperatures of \ion{H}{1} 21cm transition, which correspond to
the kinetic temperature, appear to be higher ($T \ga 1000$K) than those
typical ($T \sim 100-200$K) of the Milky Way or nearby spiral
galaxies.  Moreover, \citet{KGC01} and \citet{LBS00} found that the
large amount of neutral gas is in the warm phase in two DLA systems at
low redshifts from the spectra well fitted by multi-components with
various spin temperatures.  The inhomogeneous gas may arise from the wide
range of metallicity in present DLA systems.  Moreover, local
observations show that dusty systems generally have larger abundance.
In practice, dusty DLA systems systematically fail to be observed
because dust in DLA systems dims background quasars. 
These problems are clearly
issues that deserve further investigation.
  
In the future, the combination of accurate dust-correction using
various elements and detections of more DLA systems especially at  
$0 \la z \la 1$ are expected to reveal the origin of DLA systems 
and the star formation history at high redshifts.

\acknowledgments

 We thank M.Enoki, H.Yahagi and T.Yano for valuable discussions of 
this study and the referee for careful reading of this manuscript and 
suggestions, which improved the clarity of this presentation. 
MN also acknowledge support from a PPARC rolling grant for extragalactic
astronomy and cosmology.
This work has been supported in part by the Grant-in-Aid for Scientific 
Research Fund(13640249) of the Ministry of Education, Science, Sports and 
Culture of Japan.

\clearpage

\begin{table}
\caption{Model Parameters}  
\label{tab:astro}
\begin{tabular}{ccccccccccc}
\hline
\hline
& \multicolumn{4}{c}{cosmological parameters} &
& \multicolumn{5}{c}{astrophysical parameters} \\
\cline{2-5} \cline{7-11}
CDM Model & $\Omega_{0}$ & $\Omega_{\Lambda}$ &$h$ &$\sigma_{8}$ & &
$V_{\rm hot}$ (km~s$^{-1}$)
& $\alpha_{\rm hot}$ & $\tau_{*}^{0}$ (Gyr)
& $\alpha_*$ & $f_{\rm bulge}$ \\
\hline
LC &0.3&0.7&0.7& 1  && 280 & 2.5 & 1.5 & -2   & 0.5 \\
LD &0.3&0.7&0.7& 1  && 280 & 2.5 & 4   & -2   & 0.5 \\
SC & 1 & 0 &0.5&0.67&& 320 & 5.5 & 4   & -3.5 & 0.2 \\
OC &0.3& 0 &0.6& 1  && 220 & 4   & 1   & -3   & 0.5 \\
\hline
\end{tabular}
\end{table}

\clearpage

\begin{figure}
\plotone{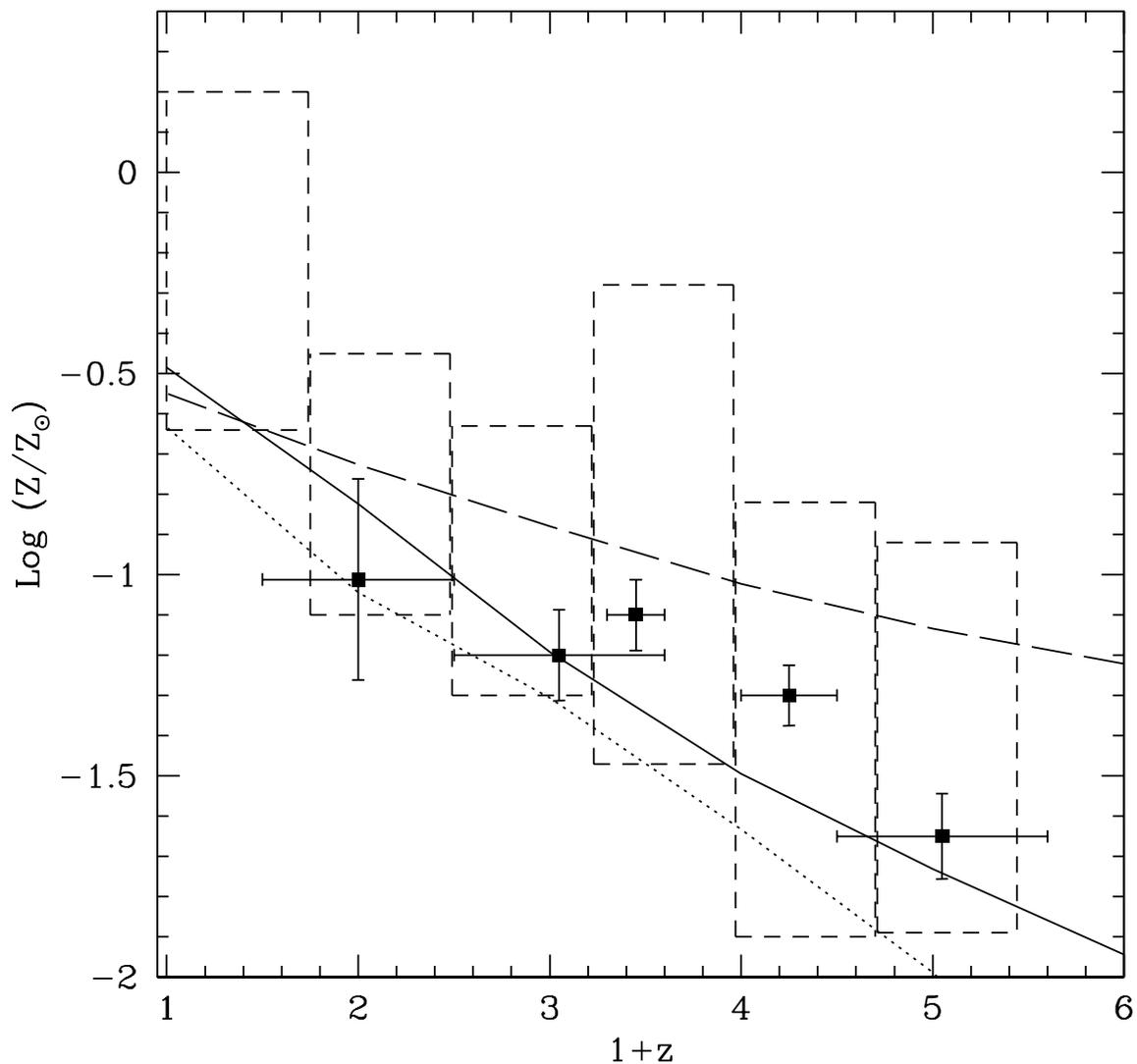}
\caption{
Metallicity of cold gas in DLA systems (in units of solar abundance) 
as a function of redshift.  The square symbols are the observed metallicity
 in DLA systems (Prochaska and Wolfe 2000). 
The boxes also represent ranges of the metallicity data corrected by dust 
depletion (Savaglio 2000).  
The solid and dashed lines  show the results of the LC  and 
LD model, respectively.  
Also shown for the metallicity evolution of 
stellar components for the LC model(dotted line). } \label{fig:corrfil_a}
\end{figure}

\begin{figure}
\plotone{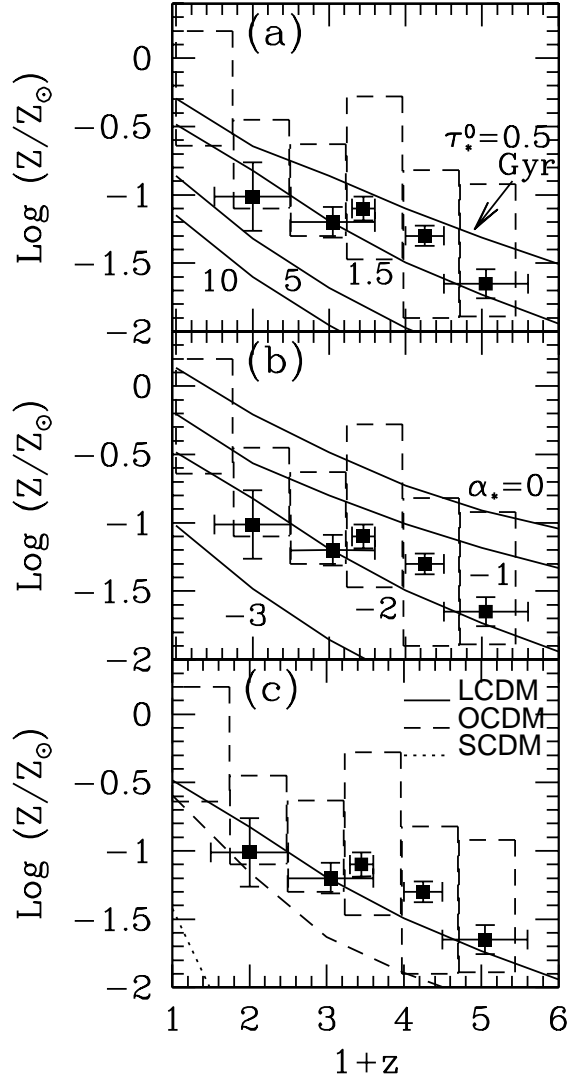}
\caption{ 
Metallicity of cold gas in DLA systems (in units of solar abundance)
 as a function of redshift.  
The observational data are also the same as in Figure 1. 
(a)The different lines depict the metallicity evolution for
various cases of star-formation timescale: 
$\tau_{*}=0.5,~1.5(Model~LC),~5 $ 
and $10~Gyr$ from top to bottom, 
(b) $\alpha_{*}=0,~-1,~-2(Model~LC)$, and $-3$ from top to bottom, 
and (c) various cosmological models: LC(solid line), OC(dashed 
line), and SC(dotted line). Cosmological parameters are summarized in Table 1.
All four cosmological models are taken from NTGY. 
} 
\label{fig:corrfil_b}
\end{figure}

\begin{figure}
\plotone{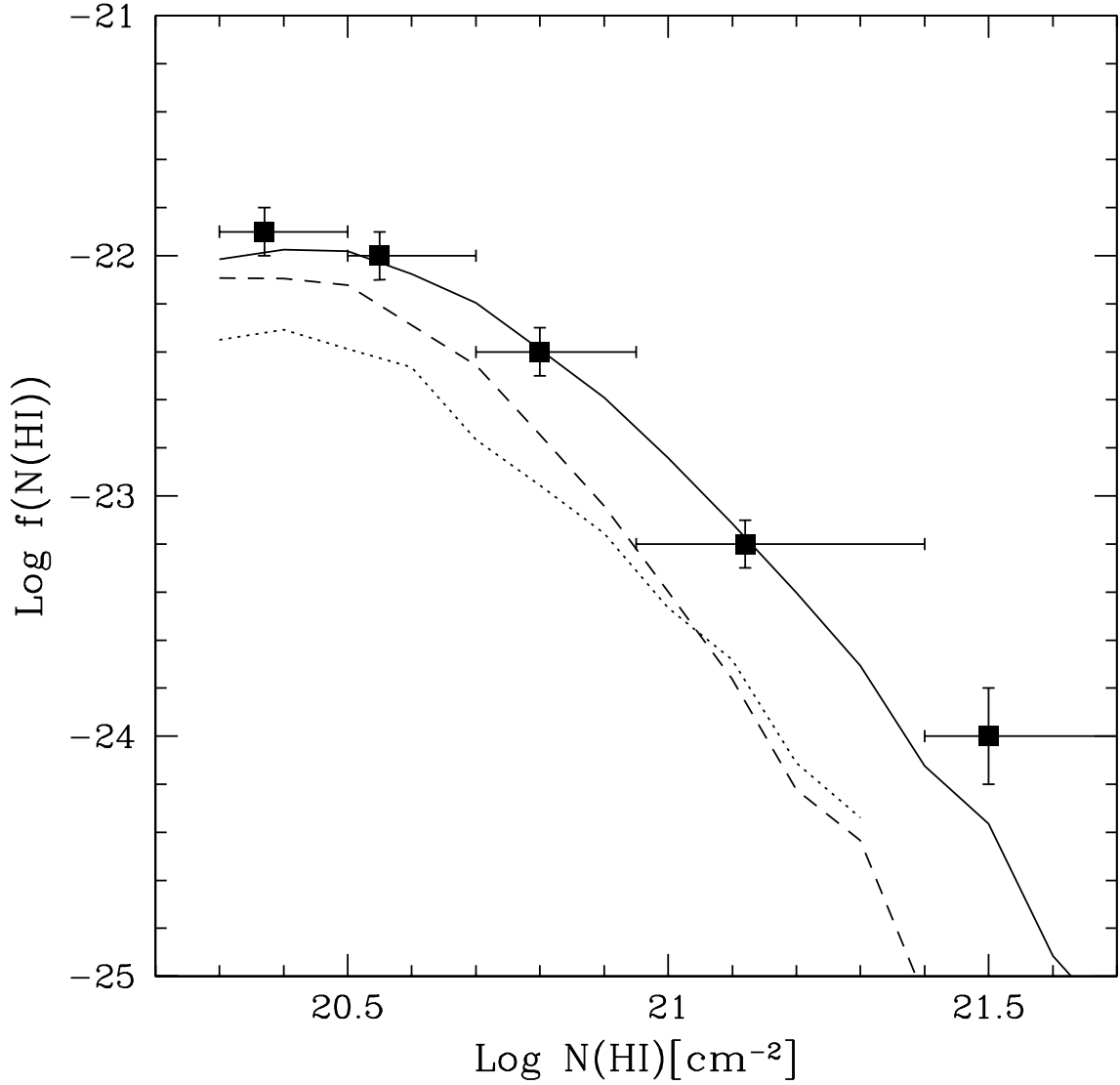}
\caption{
DLA column density distribution $f(N_{\rm HI})$ for three models: 
the LC model ($\alpha_{*}=-2$)(solid line), LC model except for 
$\alpha_{*}=0$(dotted line), and LD model(dashed line). 
 The square symbols with crosses are the observational data 
\citep{SW00}.  } \label{fig:corrfil_c}
\end{figure}

\begin{figure}
\plotone{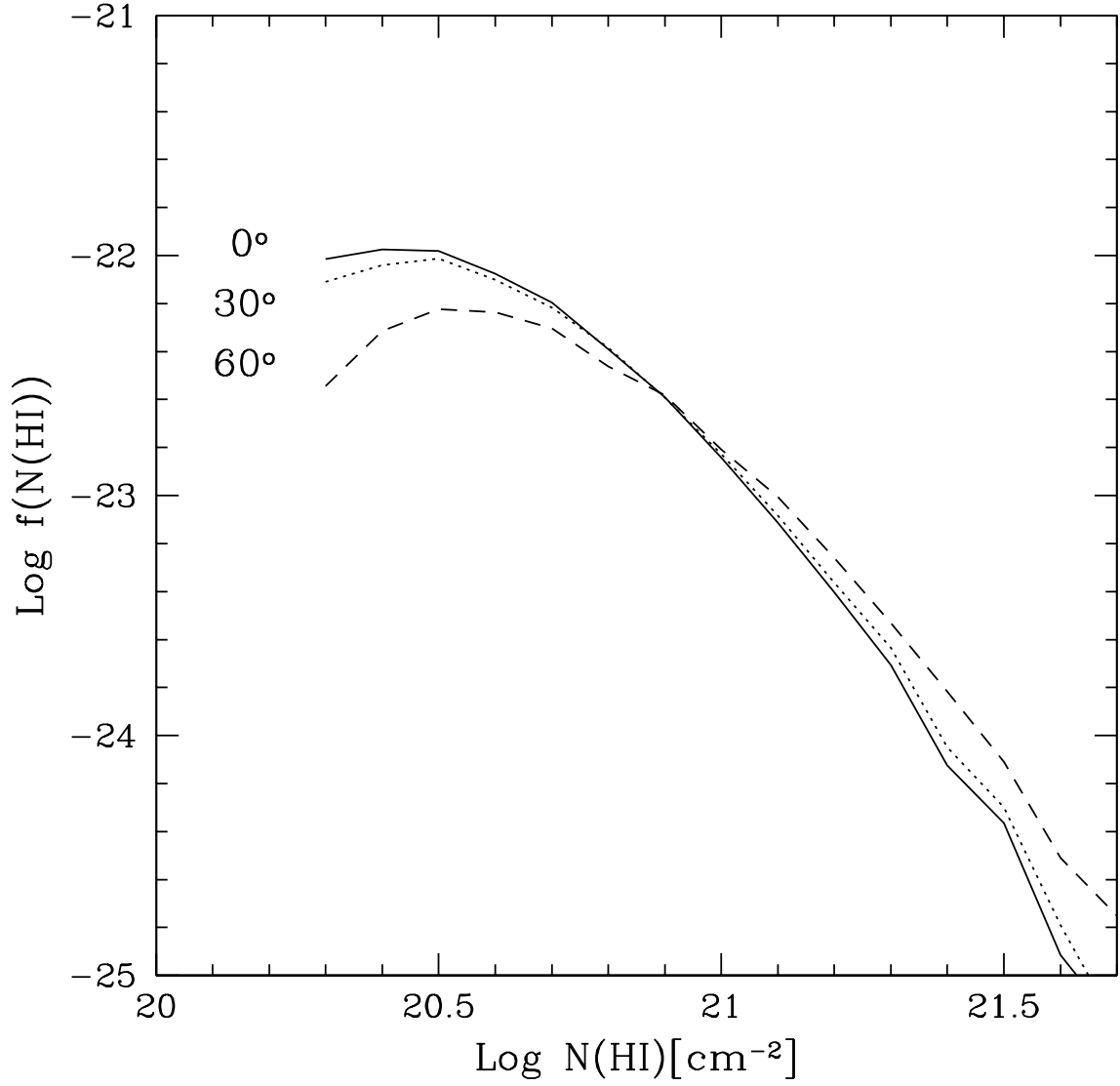}
\caption{
DLA column density distribution $f(N_{\rm HI})$ for three viewed-angles: 
$\theta=0^{\circ}$(solid line), $30^{\circ}$(dotted line), 
and $60^{\circ}$(dashed line). } \label{fig:corrfil_d}
\end{figure}

\begin{figure}
\plotone{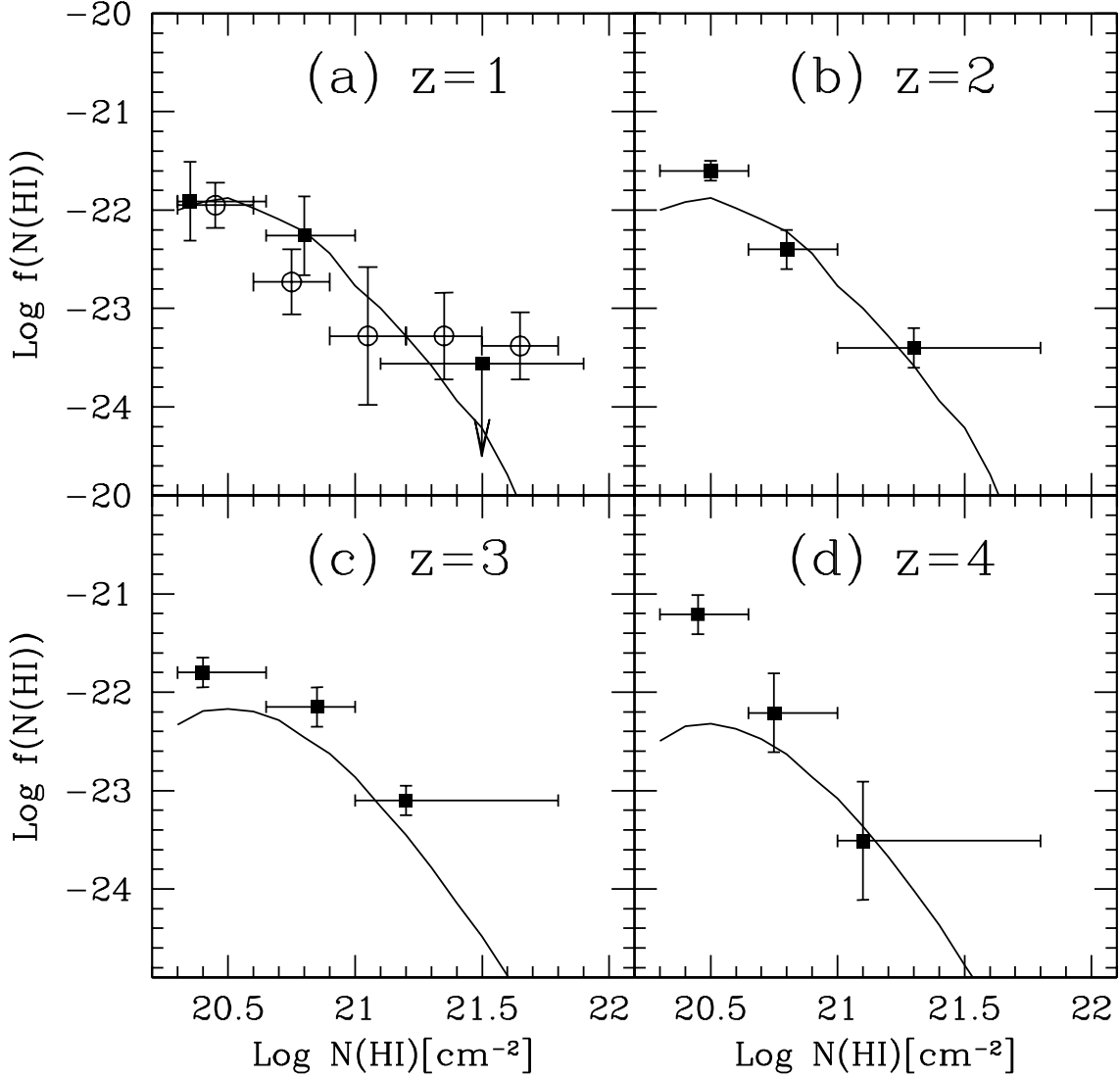}
\caption{
DLA column density distribution $f(N_{\rm HI})$ for the LC model
 at four redshifts: (a)$z=1$, (b)$z=2$, (a)$z=3$, and (d)$z=4$.
The square symbols with crosses are the observed data shown in 
four redshifts \citep{SW00}(closed square).
Also shown in Figure (a) for another observation at 
$ \langle z \rangle = 0.78$ 
\citep{RT00}(open circle).      
 } \label{fig:corrfil_e}
\end{figure}

\begin{figure}
\plotone{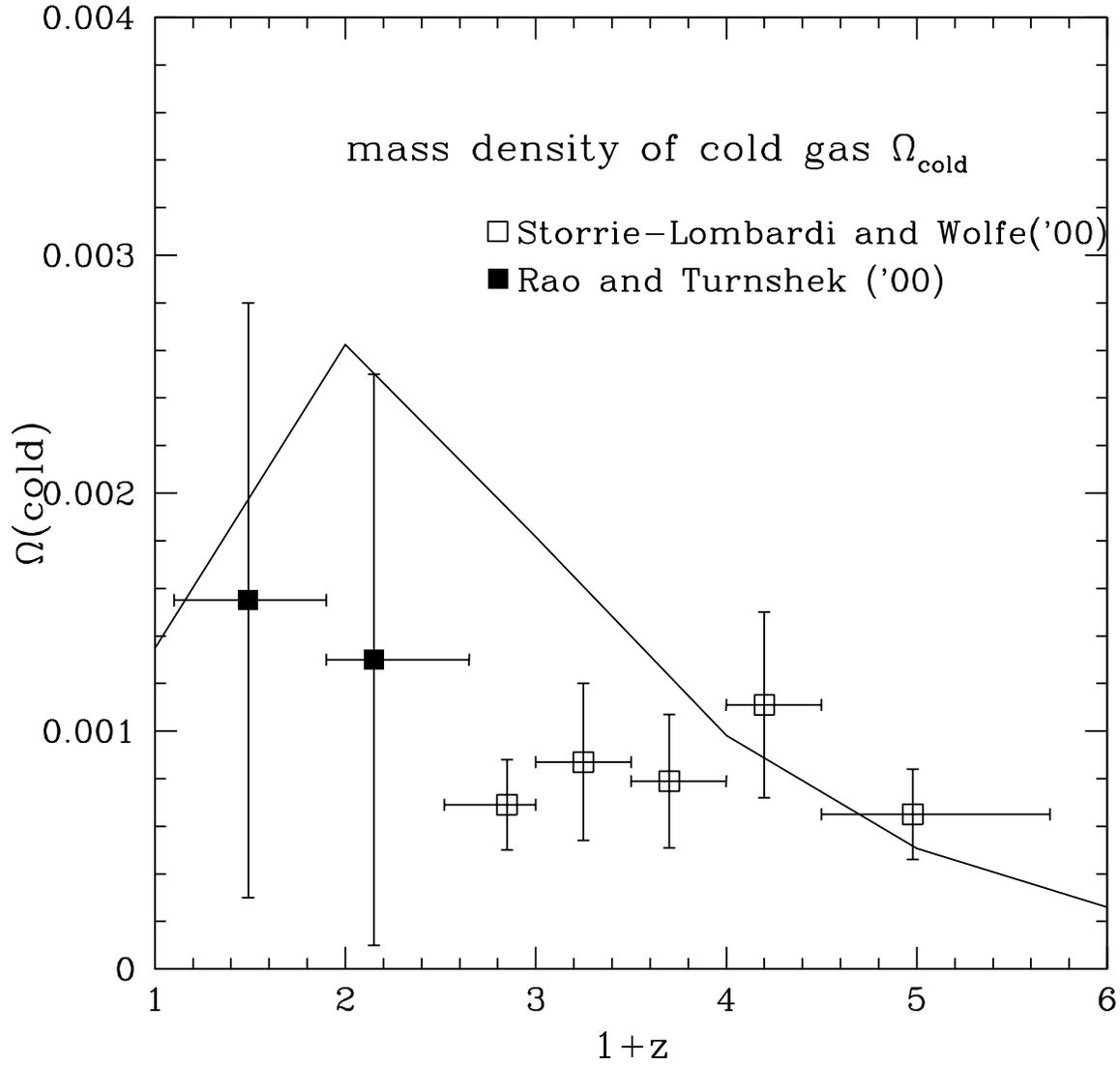}
\caption{
Mass density of cold gas in DLA systems as a function of redshifts. 
The observational data are given by \citet{SW00}(open
square) and \citet{RT00}(closed square).
} \label{fig:corrfil_f}
\end{figure}

\begin{figure}
\plotone{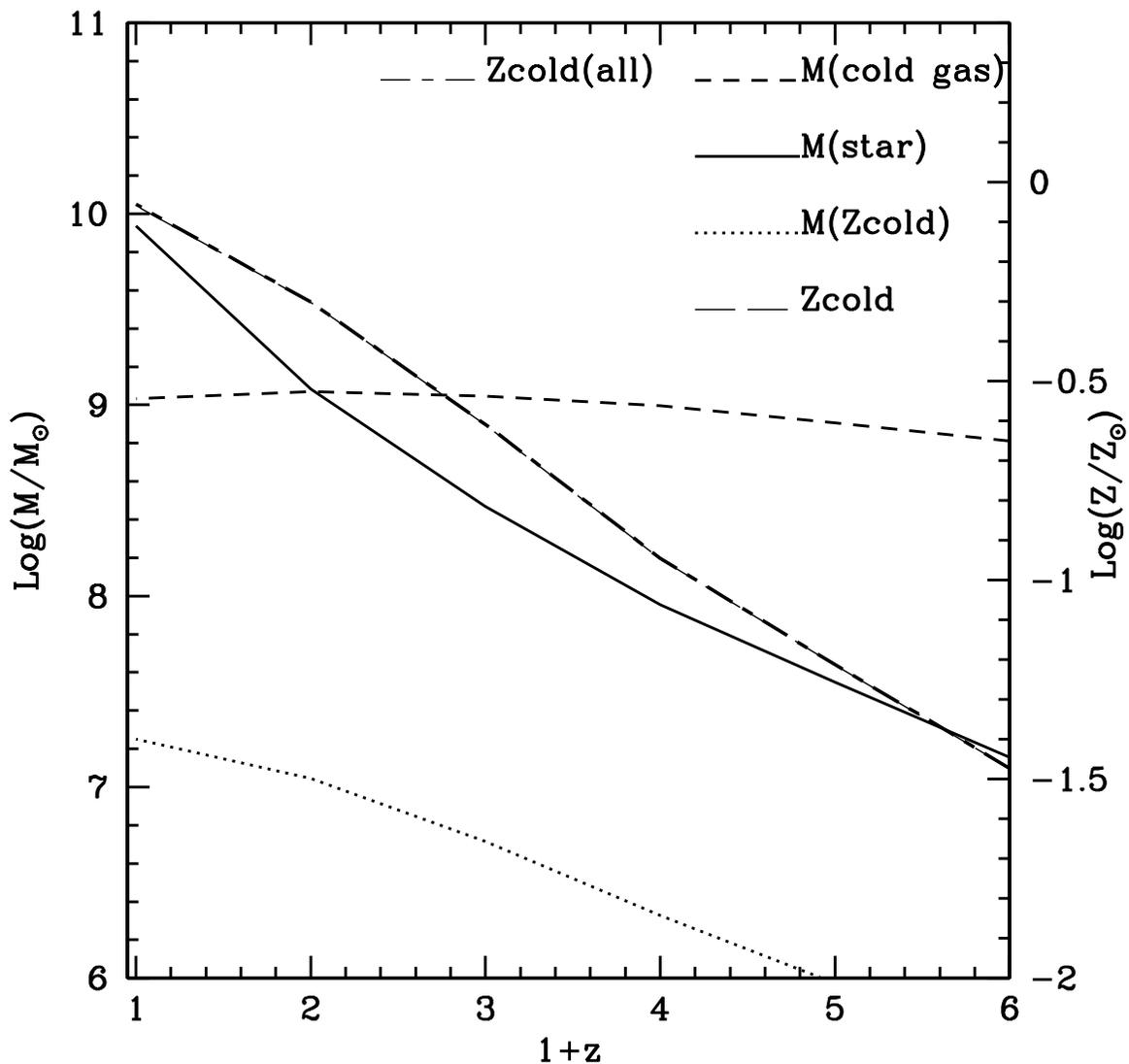}
\caption{ Average mass evolution of each phase as a 
function of redshifts:
cold gas(dashed line), stars(solid line), 
and metals(dotted line). 
Metallicity weighted by mass  $Z$cold is depicted as 
long-dashed line.  We also show the mass-weighted metallicity 
of all cold gas  $Z$cold(all) as dot-dashed line 
which is almost identical with $Z$cold(long-dashed line). 
 }
\label{fig:corrfil_g}
\end{figure}

\begin{figure}
\plotone{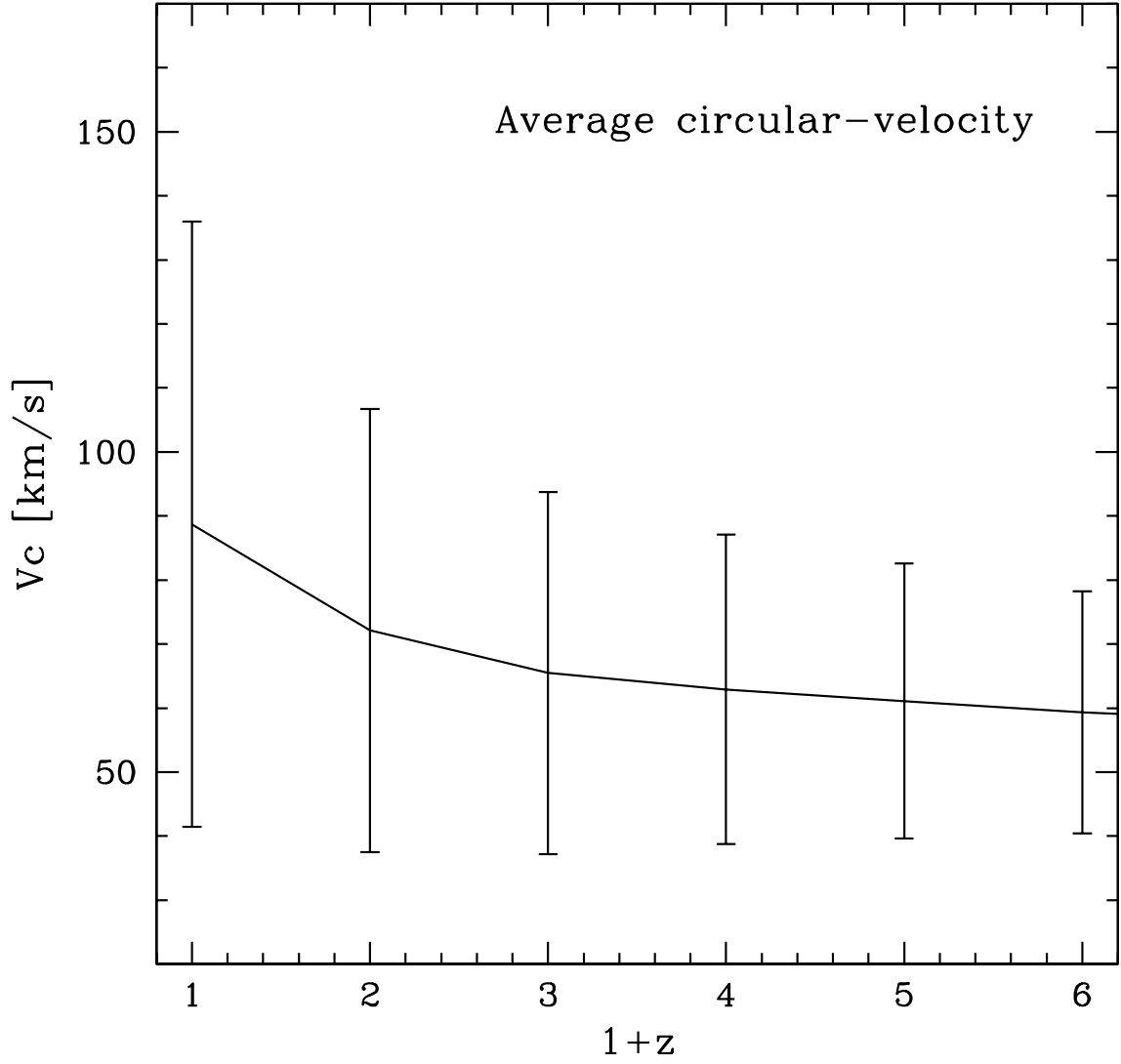}
\caption{Evolution of average circular velocity 
$V_{\rm circ}$ of DLA systems. Error bars 
with the averages indicate $1 \sigma$ errors.  
 }
\label{fig:corrfil_h}
\end{figure}

\begin{figure}
\plotone{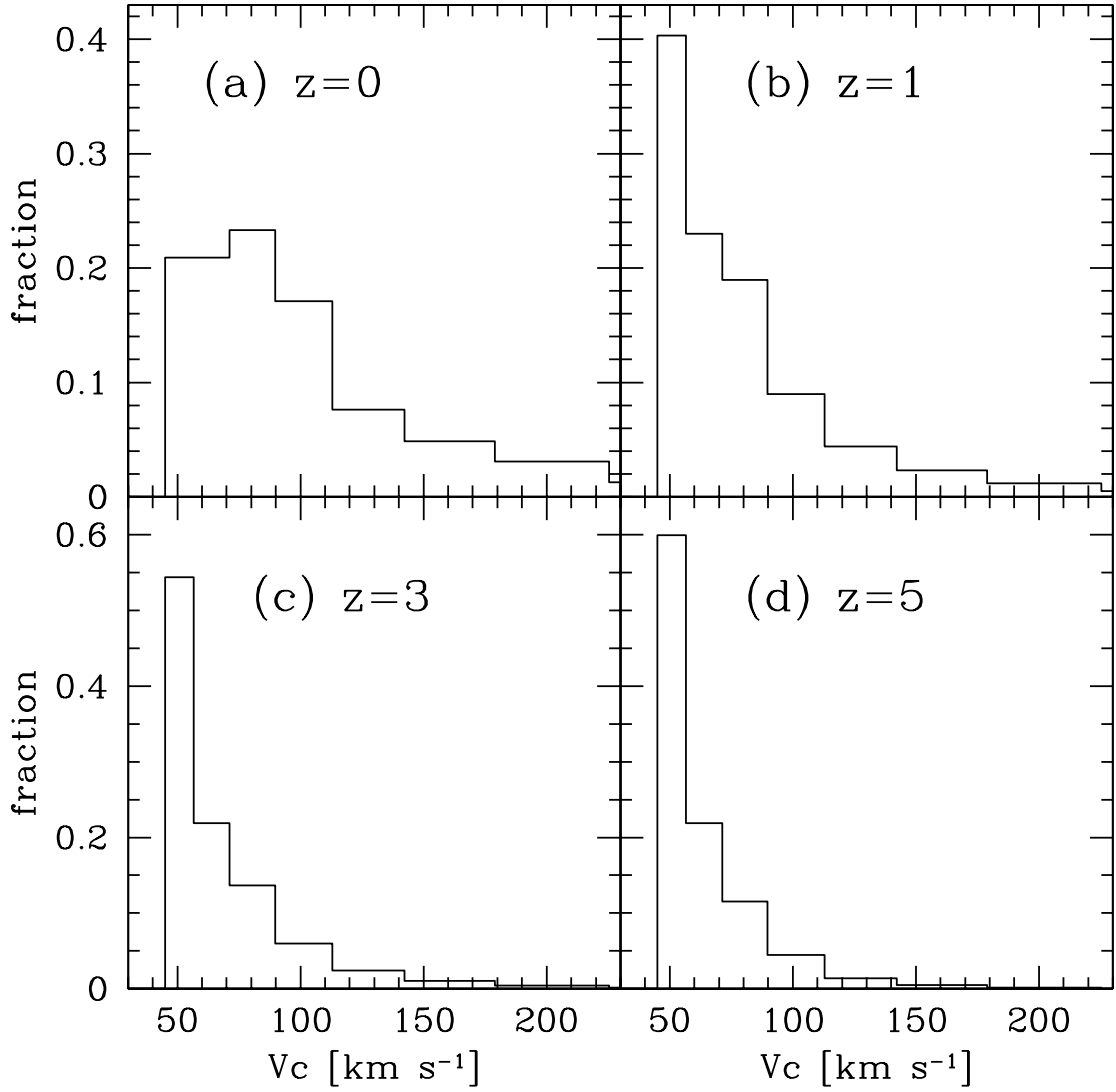}
\caption{The distribution of circular velocity $V_{\rm circ}$ 
at four redshifts: (a)$z=0$, (b)$1$, (c)$3$ and (d)$5$. 
 }
\label{fig:corrfil_i}
\end{figure}

\begin{figure}
\plotone{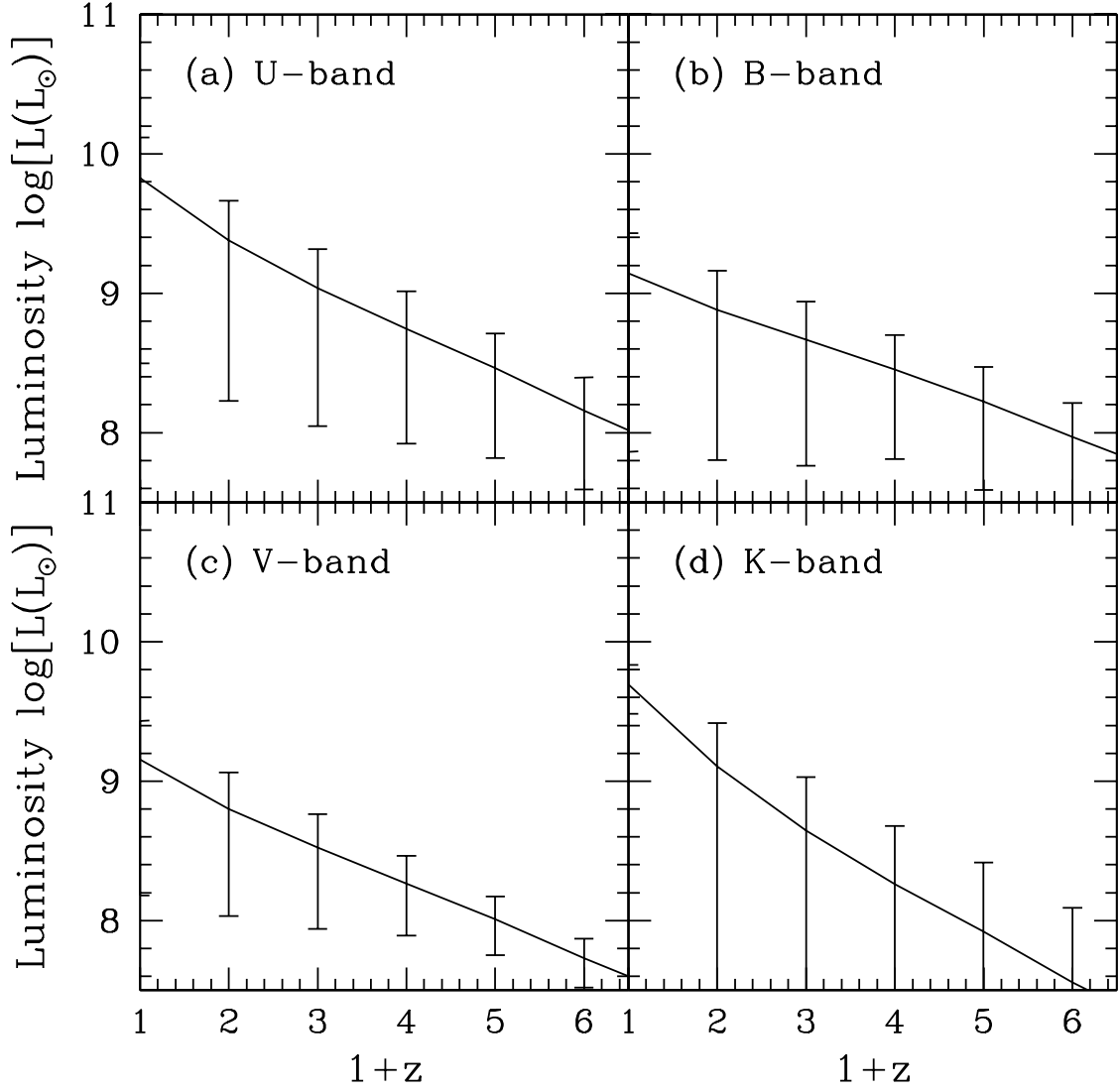}
\caption{Average luminosity evolution of host galaxies 
identified as DLA systems in four bands: 
(a)U-band, (b)B-band, (c)V-band and (d)K-band. 
$1 \sigma$ errors are shown by error bars with the averages. }
\label{fig:corrfil_j}
\end{figure}

\begin{figure}
\plotone{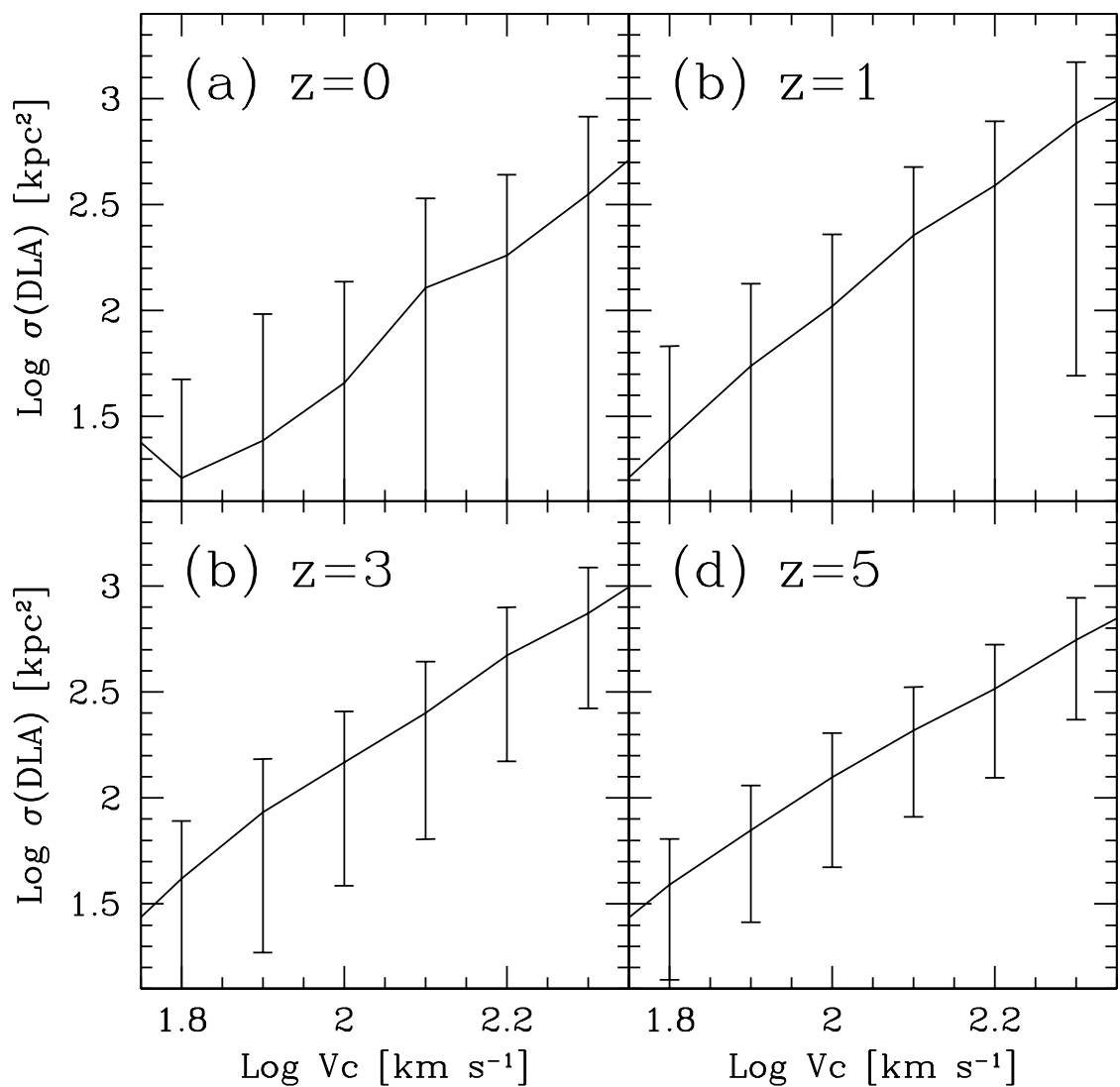}
\caption{ Average cross sections of DLA systems as a function of circular
 velocity $V_{\rm circ}$ at redshifts (a)$z=0$, (b)$1$, (c)$3$, and
 (d)$5$. In each panel, error bars indicate $1 \sigma$ errors. 
 }

\label{fig:corrfil_k}
\end{figure}

\end{document}